\documentclass [aps,prl,twocolumn,amssymb,showpacs,superscriptaddress]{revtex4-1}

\usepackage[pdftex]{graphicx}

\newcommand{\tc}{$T_c~$}
\newcommand{\tco}{$T_{c0}$~}
\newcommand{\baru}{Ba(Fe$_{1-x}$Ru$_{x}$)$_2$As$_2$~}
\newcommand{\asp}{BaFe$_2$(As$_{1-x}$P$_{x}$)$_2$~}
\newcommand{\spm}{$s_{\pm}$}
\newcommand{\spp}{$s_{++}$}
\newcommand{\ecm}{\={e}/cm$^2$}

\begin{document}

\title{Effect of electron irradiation on superconductivity in single crystals of Ba(Fe$_{1-x}$Ru$_{x}$)$_2$As$_2$ ($x=$0.24) }

\author{R.~Prozorov}
\email{Corresponding author: prozorov@ameslab.gov}
\affiliation{The Ames Laboratory, Ames, IA 50011}
\affiliation{Department of Physics \& Astronomy, Iowa State University, Ames, IA 50011}

\author{M. Ko\'{n}czykowski}
\affiliation{Laboratoire des Solides Irradi\'{e}s, CNRS-UMR 7642 \& CEA-DSM-IRAMIS, Ecole Polytechnique, F 91128 Palaiseau cedex, France}

\author{M. A. Tanatar}
\affiliation{The Ames Laboratory, Ames, IA 50011}
\affiliation{Department of Physics \& Astronomy, Iowa State University, Ames, IA 50011}

\author{A. Thaler}
\affiliation{The Ames Laboratory, Ames, IA 50011}
\affiliation{Department of Physics \& Astronomy, Iowa State University, Ames, IA 50011}

\author{S. L. Bud'ko}
\affiliation{The Ames Laboratory, Ames, IA 50011}
\affiliation{Department of Physics \& Astronomy, Iowa State University, Ames, IA 50011}

\author{P. C. Canfield}
\affiliation{The Ames Laboratory, Ames, IA 50011}
\affiliation{Department of Physics \& Astronomy, Iowa State University, Ames, IA 50011}

\author{V. Mishra }
\address{Materials Science Division, Argonne National Laboratory, Lemont, IL-60439}

\author{P. J. Hirschfeld}
\affiliation{Department of Physics, University of Florida, Gainesville, FL 32611}

\date{13 May 2014}

\begin{abstract}
A single crystal of isovalently substituted Ba(Fe$_{1-x}$Ru$_{x}$)$_2$As$_2$ ($x=0.24$) was sequentially irradiated with 2.5 MeV electrons up to a maximum dose of $2.1 \times 10^{19}$ \ecm. The electrical resistivity was measured \textit{in - situ} at $T=22$~K during the irradiation and \textit{ex - situ} as a function of temperature between  subsequent  irradiation runs. Upon irradiation, the superconducting transition temperature, \tc,  decreases and the residual resistivity,  $\rho_0$,  increases. We find that electron irradiation leads to the fastest suppression of \tc\ compared to other types of artificially introduced disorder, probably due to the strong short-range potential of the point-like irradiation defects. A more detailed analysis within a multiband scenario with variable scattering potential strength shows that the observed \tc\ vs. $\rho_0$ is fully compatible with  \spm\ pairing, in contrast to earlier claims that this model leads to a too rapid a suppression of \tc\ with scattering.
\end{abstract}



\maketitle

There are several experimental approaches to study the superconducting gap structure in superconductors. One of them is to measure the change of the superconducting transition temperature, \tc, with artificially introduced disorder. Since impurity scattering mixes the superconducting order parameter at different points on the Fermi surface, controlled potential disorder may be considered a phase-sensitive probe of gap symmetry. It is well known that while the gap and critical temperature of an isotropic $s-$wave superconductor are insensitive to nonmagnetic disorder \cite{Anderson1959a,Abrikosov1960}, superconducting states with different gap symmetries and structures may be extremely sensitive \cite{BalianWerthamer1963,Openov1998,KontaniJPSJ2008,Kontani2009,EfremovPRB2011,TcEnhancement2012,Wang2013}. In case of iron-based superconductors, the predictions for the effect of disorder differ for various possible pairing states and depend on details of the model. In particular,  models involving repulsive interactions, including popular spin fluctuation models (for a review see Ref.~ \onlinecite{HKM_review}) predict states where the order parameter changes signs between sheets of the Fermi surface, generically called $s_\pm$ here, whereas models  involving orbital fluctuations \cite{KontaniJPSJ2008,Kontani2009} and attractive interactions  predict no sign change ($s_{++}$). The effect of disorder has also been studied in the coexisting superconducting and long-range magnetic order phase \cite{TcEnhancement2012}. These different approaches can be studied within a phenomenological multiband theory that for some parameters predicts a crossover from the \spm\ to the \spp state \cite{EfremovPRB2011}.

The major experimental problem in the studies of the effect of disorder is the actual introduction of point-like defects with a minimal impact on the material itself. In case of chemical substitutions, there is always a question of whether the foreign ions change not only the scattering, but other parameters such as chemical potential and the band-structure. These studies revealed ``slow" $T_c$ suppression in 122 systems, which was interpreted as the evidence for $s_{++}$ pairing \cite{Lietal2012}. Recently Wang \textit{et al.} criticized this conclusion by  extending the phenomenological multiband impurity scattering model to include gap anisotropy, and by exploring the effect of differing ratios of intra- to interband scattering matrix elements \cite{Wang2013}. They showed that the rate of $T_c$ suppression depended sensitively on this ratio, and  argued that the relatively slow suppression of $T_c$ in some chemically doped Fe-based materials was due to dominant intraband scattering. Indeed, it is clear that different ions result in very different scattering mechanisms and may  also in some cases induce a magnetic moment, which, of course, changes the scattering and pair breaking rates \cite{ChengMnZn2010,Konbu2012}. With this perspective, it is  not so surprising that different studies even for the same impurity ion, for example Zn, show completely different results \cite{ChengMnZn2010,ZnDopingPRB2011,Li2012}.

Irradiation with energetic particles is an alternative way to introduce  defects. However, the nature of the produced defects depends on the type of irradiation \cite{Damask1963}. Heavy - ion irradiation produces columnar tracks or sausage-like linear defects \cite{Nakajima2009}, which are difficult to analyze within simplified point-like potential scattering models. Yet, the experiments with heavy ion irradiation in iron pnictides have shown a definitive violation of the Anderson theorem \cite{Nakajima2009,Kim2010,Prozorov2010a}, and independent measurements of the London penetration depth and \tc\ in BaCo122 and BaNi122 allowed the elimination of the unknown scattering rate, with  the analysis then favoring the \spm\ pairing scenario \cite{Kim2010,Prozorov2010a,MurphyFeCo2013}. Proton \cite{Fang2011,Nakajima2010,Taen2013} and $\alpha -$ particle \cite{Tarantini2010} irradiations were also used to study iron pnictides. Proton irradiation has thus far produced greatest $T_c$ suppression rate \cite{Taen2013}, albeit 2 - 7 times slower than reported here for electron irradiation. Analysis of the energy transfer from an accelerated particle smashing into the crystal lattice shows that only electrons with energies of  1 to 10 MeV produce point - like defects in form of interstitial ions and vacancies (Frankel pairs) that presumably form perfect scattering centers \cite{Damask1963}. Indeed, these defects are charged, but the overall charge change is compensated, so that there is a negligible shift of the chemical potential due to irradiation. Protons, $\alpha-$ particles and neutrons most likely produce cascades of clusters of defects, and heavy ions produce columnar tracks. A more detailed and systematic investigation of the connection between the type of the scattering centers (their spatial extent and scattering strength) and pair-breaking is needed. Thus far, the effect of the finite size of the defects on $T_c$ suppression rate was studied theoretically in Ref.~\onlinecite{Yamakawa2013}. The effect of electron irradiation on electron - doped BaCo122 and BaNi122 was compared with the effect on Ba(AsP)122 in Ref.~\cite{vanderBeek2013} and it appears that isovalent systems are closer to a clean limit than charge - doped ones.

Isovalent to iron, ruthenium substitution suppresses long - range magnetic order and induces superconductivity, but does not change the compensation condition between holes and electrons \cite{Rullier-Albenque2010,Thaler2010,Shishido2010,Dhaka2011,Xu2012}. Angle - resolved photoemission spectroscopy (ARPES) measurements report no change in the shape of the Fermi surface and Fermi velocities up to $x=0.55$ \cite{Dhaka2011} while other ARPES study finds a crossover from two- to three- dimensional geometry of some hole - like pockets of the Fermi surface accompanied by the mass increase for the larger doping levels $x$ \cite{Xu2012}. These results were theoretically analyzed in a recent \textit{ab initio} study \cite{LWang2013}. For comparison, in another isovalently substituted system, \asp, Fermi surface shrinks with $x$ decreasing from $x=1$ \cite{Shishido2010} and the effective mass diverges on the approach of the optimal doping at $x=0.3$ consistent with the quantum critical point beneath the dome \cite{AsP122Science2012}. For the purpose of this work, it is important that for the doping level discussed here, $x=0.24$, the Ru substitution causes no carriers imbalance.

The discussion of the effect of the disorder in iron pnictides crucially depends on the multiband nature of superconductivity that supports both $s_\pm$ and $s_{++}$ pairing states \cite{EfremovPRB2011}. It is therefore important to compare the effects of irradiation in iron pnictides with an established two-gap $s_{++}$ superconductor, $MgB_2$. As expected from the Anderson theorem, both low dose \cite{Blinkin2006} and higher dose \cite{KleinPRL2010MgB2} (comparable to this study) electron irradiations found virtually no change of \tc clearly supporting $s_{++}$ nature of superconductivity. On the other hand, neutron irradiation $MgB_2$ led to a complete suppression of $T_c$ \cite{Wilke2006}, which prompts the question on the nature of defects produced by neutron irradiation with a possibility of a nuclear transmutation of boron into carbon. 

In this paper, we report \textit{in-situ} and \textit{ex-situ} measurements of the electrical resistivity in a single crystal of isovalently substituted \baru\ ($x=$0.24) subsequently irradiated with 2.5 MeV electrons of different doses up to $2.1 \times 10^{19}$ \ecm. To avoid ambiguity in determination of the scattering rate, we exhibit the change of \tc\ versus measured residual resistivity $\rho_0$ and we also calculate the conventional single scattering rate, $g^{\lambda}$ to compare our results with the previous studies and theoretical predictions. Our results indicate that in this system \tc\ is suppressed very rapidly by point - like potential scattering.  Since this is not possible in $s_{++}$ superconductors, our results  provide strong evidence for the \spm\ pairing mechanism in iron - based systems.

\begin{figure}[tb]%
\includegraphics[width=8cm]{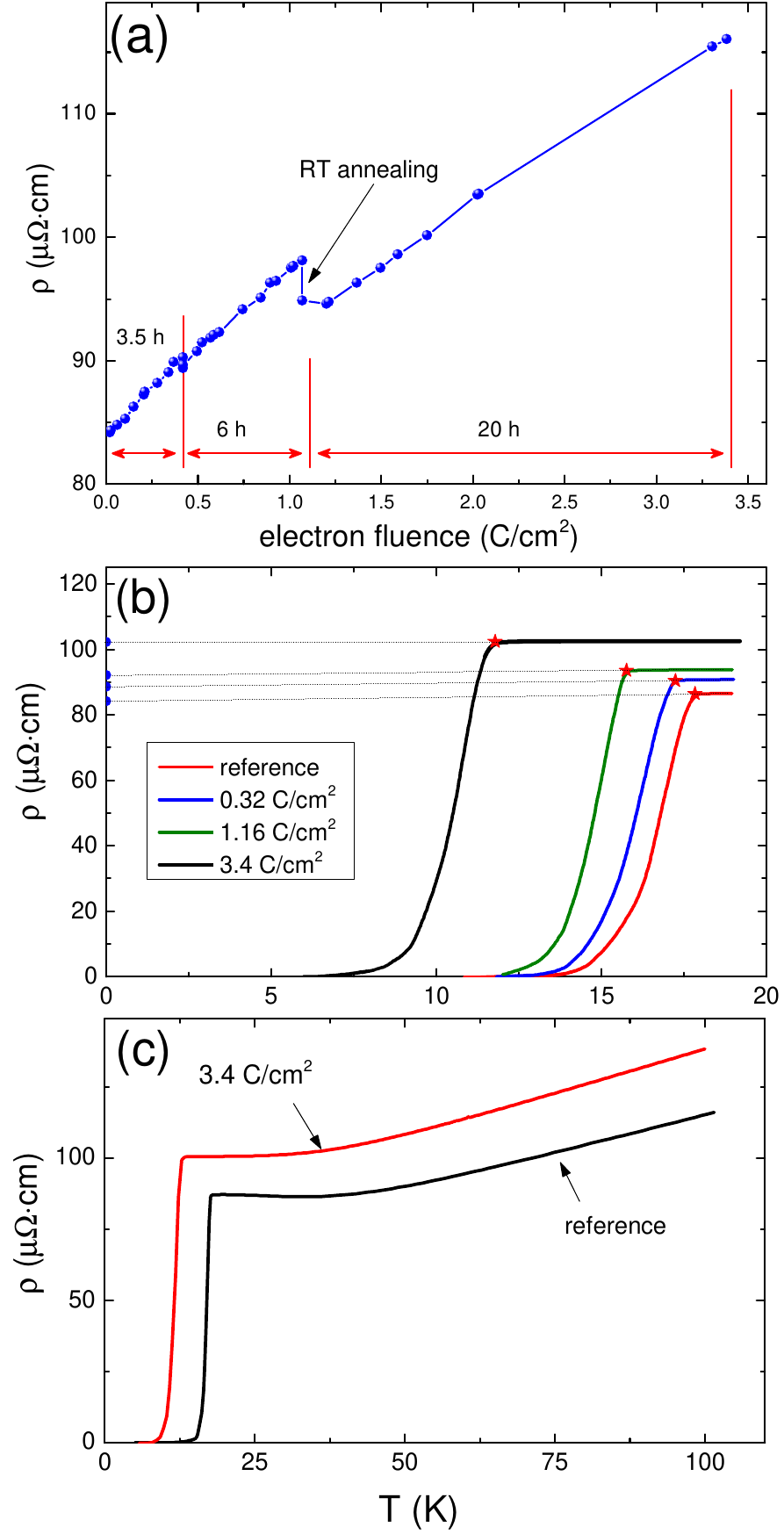}%
\caption{(Color online) (a) The \textit{in - situ} measurements of resistivity in \baru\ ($x=$0.24) at $T \approx 22$K as a function of the irradiation dose. The breaks in the curve correspond to the extraction of the sample and warming it up to the room temperature resulting in a partial annealing of the defects. (b) The \textit{ex - situ} measurements of resistivity vs. temperature between the irradiation runs. (c) Extended temperature range resistivity showing practically uniform shift of the entire curve upon irradiation.}
\label{fig1}%
\end{figure}


{\it Samples and irradiation technique.} Single crystals of \baru\ ($x=0.24$, $T_{c0} = 17.8$~K), were grown out of self-flux as described in detail in Ref.~\onlinecite{Thaler2010}. The samples were characterized with x-ray diffraction, magnetization, transport and magneto-optical measurements. The composition was determined by using wavelength dispersive x-ray spectroscopy (WDS) in a JEOL JXA-8200 electron microprobe. The 2.5 MeV electron irradiation was performed at the SIRIUS Pelletron linear accelerator operated by the \textit{Laboratoire des Solides Irradi\'{e}s} (LSI) at the \textit{Ecole Polytechnique} in Palaiseau, France. The sample with four soldered contacts \cite{Tanatar2010a} was mounted inside the cell with a flow of liquid hydrogen at $T \approx 22$~K. 
The \textit{in-situ} electric transport measurements were performed while irradiating the sample with the electron beam of 5 to 8 $\mu$A total current through a 5 mm diameter diaphragm. This current (which eventually provides the calibration of the irradiation dose) was measured with the Faraday cage placed behind the sample stage. The irradiation rate was about $3\times10^{-5}~\mathrm{C}/\mathrm{s}~ \mathrm{cm}^2$ and the experiments lasted several days. For convenience, we express the irradiation dose in C/cm$^2$. The actual dose in the number of electrons per cm$^2$ can be obtained by dividing this number by the electron charge \=e. 

{\it Results.} Figure \ref{fig1}(a) shows \textit{in - situ} measurements of four-probe resistivity at 22 K in BaRu122 single crystal during electron irradiation. The irradiation was stopped twice, sample warmed up to room temperature, transferred to another cryostat in which $\rho(T)$ was measured and the sample returned to the irradiation chamber. The contacts were never altered in the process. Warming up to the room temperature results in a partial annealing of the induced defects. By analyzing the experiments with different samples and doses, a conservative estimate of the annealing rate is about 30 \% maximum, after which the defects find a metastable configuration and remain localized. This was verified by re-measuring the same samples after months of storage at room temperature in a desiccator. Figure \ref{fig1}(b) shows temperature - dependent resistivity, $\rho(T)$, measured between the irradiation runs in a separate cryostat. Clearly, the resistivity increases and the transition temperature decrease with irradiation.  In Fig.~\ref{fig1}(c) we show further that to a good approximation, the resistivity increase $\Delta\rho_0$ is practically independent of temperature.

\begin{figure}[tb]%
\includegraphics[width=8cm]{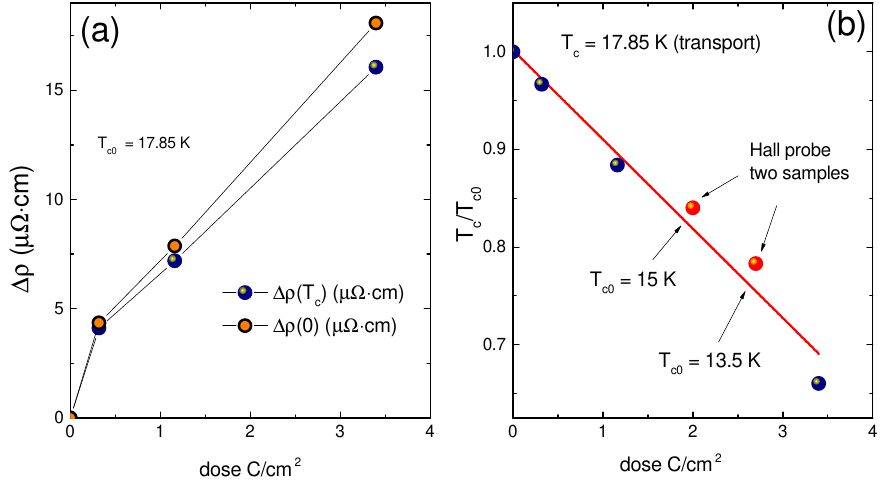}%
\caption{(Color online) (a) Change of resistivity of \baru\ ($x=$0.24) at \tc\ and extrapolated to $T=$0 (see Fig.~\ref{fig1}(b)) as function of the irradiation dose. (b) Suppression of \tc\ with the dose of electron irradiation.}
\label{fig2}%
\end{figure}

Figure \ref{fig2}(a) shows the increase of four-probe in-plane resistivity at \tc\ and extrapolated to $T=$0 (see Fig.~\ref{fig1}(b)) as a function of the irradiation dose, and Fig.~\ref{fig2}(b) shows the suppression of \tc\ with the electron irradiation dose. Two additional points are from two other samples of BaRu122 with quite different initial \tco\ and irradiated at the indicated doses. Apparently they fall onto a universal curve indicating that the rate of suppression of $T_c$ is independent of the composition in this material as expected for the isovalent substitution. If the irradiation had another effect in addition to the introduction of disorder (e.g., shift of the chemical potential), such universality would unlikely be observed. This is also confirmed by the direct Hall effect measurements in which we found no practically change of the Hall coefficient.

\begin{figure}[tb]%
\includegraphics[width=8 cm]{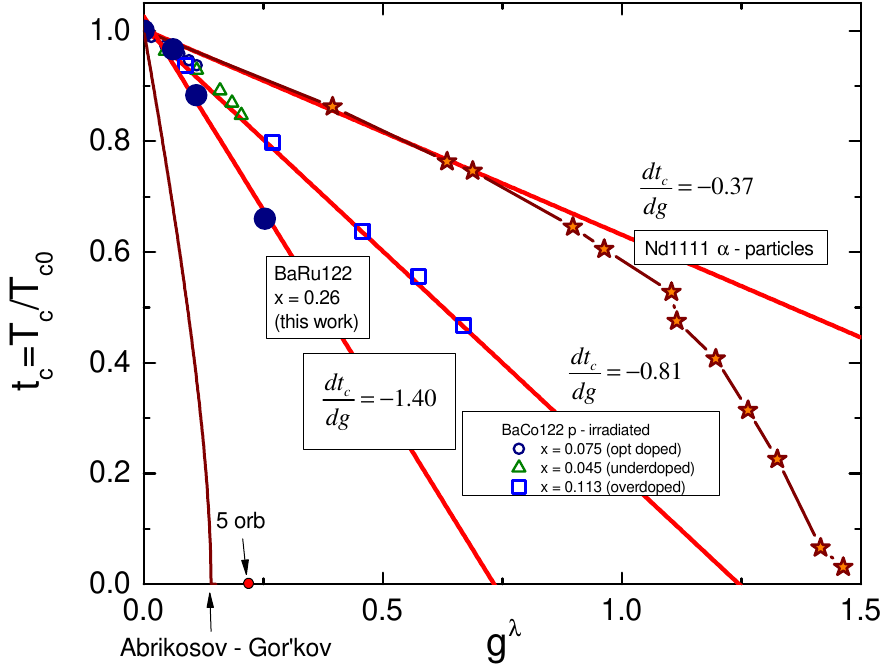}%
\caption{(Color online) Comparison of the \tc\ suppression by three irradiation techniques used to introduce artificial disorder in iron - pnictides. The single effective dimensionless scattering rate, $g^{\lambda}$, was calculated from the penetration depth and resistivity, see text for description. Abrikosov - Gor'kov theory for an isotropic $s-$wave superconductor with magnetic impurities (solid line) and a critical scattering rate within 5 band $s_\pm$ model \citep{Kontani2009} are also shown.}
\label{fig3}%
\end{figure}

Figure \ref{fig3} compares our results with previous studies of the effect of artificial disorder. Proton irradiation was used on FeCo122 crystals of three different doping levels \cite{Nakajima2010},  $\alpha-$ particle were used to irradiate a very thin Nd1111 crystal \cite{Tarantini2010}. The difference in the rate of \tc\ suppression is consistent with the energy - loss calculations predicting that $\alpha-$particles produce more correlated clusters, protons still produce clusters and the electrons produce point - like defects, which are most efficient pairbreaking scattering centers due to the localized nature of the scattering potential. The analysis presented in these studies, however, used a single dimensionless scattering rate, $g$, and was based on the original prediction of the \spm\ model \cite{Kontani2009,Onari2010}. The Authors concluded that the rate of change in $T_c(g)$ is too slow for the \spm\ scenario. In our view, this conclusion has several flaws. First is the assumption of a single scattering rate $g$ for a single type of defects, although these types of irradiation tend to create a distribution of defects.  Secondly, in these works $g$ is estimated within an isotropic Fermi  gas model where only mass is renormalized, and assumed to represent only interband scattering. The third problem is the treatment of the scattering in the \spm\ scenario,  where equal densities of states and equal gap magnitudes on two bands are assumed (we refer to this as the symmetric model).  This set of assumptions indeed produces \tc suppression at a rate identical to the Abrikosov-Gor'kov rate \cite{Abrikosov1960}, but relaxing any of them produces a much slower relaxation rate in an $s_\pm$ state \cite{Wang2013}.

In the simplest form, the dimensionless scattering rate can be defined using the Drude model as \cite{Kogan2009},

\begin{equation}
 g^{\lambda }=\frac{\hbar }{2\pi k_{B}\mu _{0}}\frac{\Delta \rho _{0}}{T_{c0}\lambda ^{2}}\approx 0.24\frac{\Delta \rho _{0}[\mu \Omega \cdot \text{cm}]}{T_{c0}[\text{K}]}
\label{eq:gamma_lambda}
\end{equation}

\noindent where $\Delta \rho _{0}$ is the change in the residual resistivity due to irradiation and we used London penetration depth, $\lambda =$ 200 nm \cite{Gordon2010}.
This method provides a meaningful estimate of the single scattering rate related to directly measurable quantities \cite{Kogan2009}. We used $g^{\lambda} = 0.25 \Delta \rho _{0}/T_{c0}$ to plot the data in Fig.~\ref{fig3}.

{\it Discussion.} The residual resistivity change, $\Delta \rho_0$, induced by irradiation is the most useful measure of the scattering since the rate $g$ itself is not directly measurable. To avoid the ambiguity in evaluating the generalized scattering rate \cite{EfremovPRB2011},  we will use the same set of parameters to calculate \tc\ and the residual resistivity in the Drude approximation \cite{Wang2013}.  As shown in Fig.~\ref{fig2}, the increase of the resistivity induced by the electron irradiation is practically $T-$independent, meaning we can ignore the interference processes between inelastic and elastic scattering. We therefore  calculate $\Delta \rho_0$  in the same $t$-matrix  framework used in earlier studies (see for example Refs.~\onlinecite{EfremovPRB2011,Wang2013}), assuming a two band model.  We further take all the defects to be point - like and scatter within a given band with the potential $v$ (intraband), and between the bands with the potential $u$ (interband). The ratio of inter- to intraband scattering, $\alpha\equiv u/v$, is a crucial parameter determining the rate of $T_c$ suppression \cite{EfremovPRB2011,Wang2013}.

Near $T_c$, the equations for the self energies and superconducting gap can be linearized. To further simplify and avoid multiple free parameters, we take equal densities of states $N_0$ for both bands and the same intraband scattering strength $v$ and interband scattering $u$.  Now we can write for $t_c=T_c/T_{c0}$,

\begin{eqnarray}
 \log t_c = \Psi\left(\frac{1}{2}\right) -\Psi\left(\frac{1}{2}+\frac{2g_p}{t_c}\right).
\end{eqnarray}

Here $g_p$ is the pairbreaking parameter which reads
\begin{eqnarray}
 g_p &=& \frac{1}{4\pi \tau T_{c0}} \frac{\alpha^2}{\left(1+\eta^2 (1+\alpha^2)\right)^2-4\alpha^2 \eta^2},
\end{eqnarray}

\noindent here $\tau^{-1}= 2 n_{imp} \pi N_0 v^2 $ and $\eta=\pi N_0 v$ and $N_0$ is the total density of states. The unitary limit is achieved by taking $\eta\rightarrow \infty$.

The total dc conductivity in the $x-$direction is the sum of the Drude  conductivities of the two bands, $\sigma=\sigma_a+\sigma_b,$ with $\sigma_\alpha=2e^2N_{\alpha} \langle v_{\alpha,x}^2\rangle \tau_\alpha$, where $v_{\alpha,x}$ is the component of the Fermi velocity in the $x$-direction and $\tau_\alpha$ is the corresponding single particle relaxation time obtained from the self-energy in the $t-$ matrix approximation,  $\tau_\alpha^{-1}=-2\mathop{\mathrm {Im}}\Sigma_{\alpha}^{(0)}$,   containing  contributions from both intraband and interband impurity scattering processes.  The transport time is the same as the single-particle lifetime within this model because of our assumption of point - like $s-$wave scatterers, meaning that the vertex corrections vanish in the limit where the Fermi surface pockets corresponding to the two bands are small.

\begin{figure}[tb]%
\includegraphics[width=8cm]{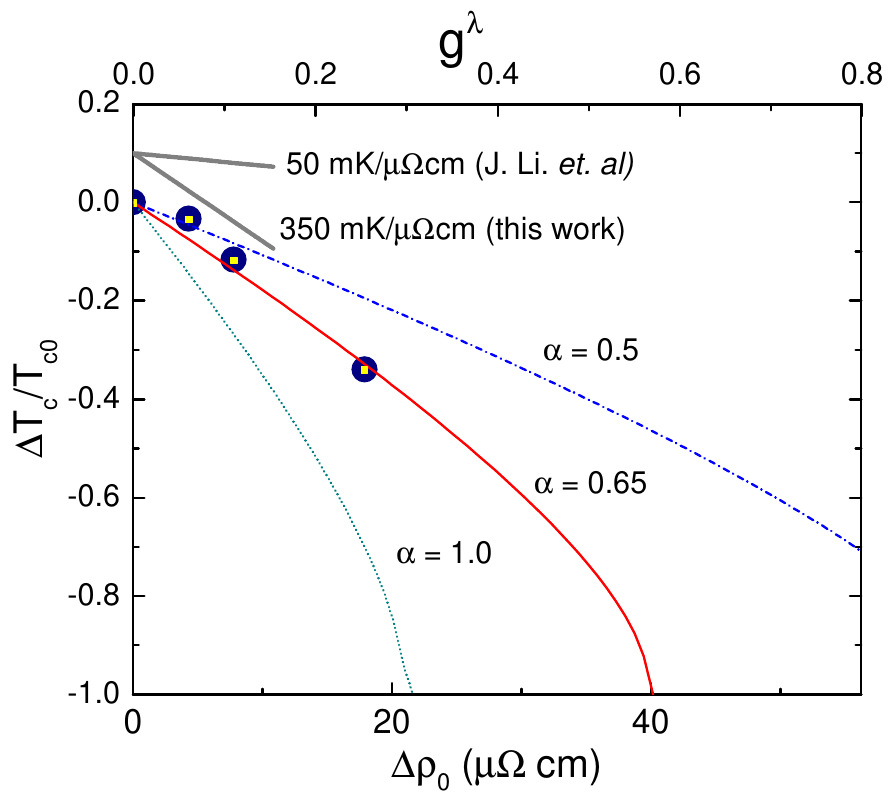}%
\caption{(Color online) Experimental data for $\Delta T_c/T_{c0}$ (symbols) vs. measured change in resistivity $\Delta \rho_0$ for single crystal of \baru\ ($x=$0.24). Lines are the fits to a $t-$matrix calculation for disorder self-energy with a fixed $\eta=0.74$, assuming the same density of states and taking Fermi velocity of $v_F=2\times 10^6$ m/s for both bands. Three theory lines correspond to the ratio of interband to intraband scattering $\alpha =$ 1.0, 0.65 and 0.5. Solid straight lines show average $T_c$ suppression of 50 $mK/\mu \Omega \cdot cm$ by transition metal impurities M substituted into Ba$_{0.5}$K$_{0.5}$Fe$_{2−2x}$M$_{2x}$As$_2$ from Ref.~\onlinecite{Lietal2012} and obtained in this work 350 $mK/\mu \Omega \cdot cm$. The top axis shows dimentionless scattering rate $g^{\lambda}$ calculated from the resistivity and penetration depth, Eq.~\ref{eq:gamma_lambda}.}
\label{fig4}%
\end{figure}

Figure~\ref{fig4} shows the variation of $T_c/T_{c0}$ with residual resistivity for three possible values of the interband and intraband potentials. Our aim here is not to fix these values, - this is not possible since there are three independent parameters $\tau^{-1}$, $u$, and $v$. However we show that fits to the data can reproduce any experimental rate of the suppression of the $T_c$ by disorder. The experimental data shown by symbols agree quite reasonably with the theory.  Note that we have excluded any possible anisotropy in order parameters so far, because determining order parameter anisotropy based only on $T_c$ suppression is not possible. In addition, the results reported in Ref.~\onlinecite{Wang2013} explicitly showed that the gap anisotropy does not affect the qualitative features of the $T_c$ suppression problem.

On the same plot, we show a line indicating the $T_c$ suppression found by Li \textit{et al.} \cite{Lietal2012} for a variety of different transition metal impurities (including magnetic Mn ions) all of which suppressed $T_c$ at roughly the same rate, a factor of 7 slower than that caused by the irradiation in our study. We note, however, that another isovalently substituted system, BaFe$_2$(As$_{1-x}$P$_x$)$_2$, at the optimal doping shows remarkably close rate of 340 $mK/\mu \Omega \cdot cm$. We also note that the pure sample used by Li \textit{et al.} had initial residual resistivity significantly higher than the unirradiated sample used in our study.  Nevertheless, the remarkable difference between the $T_c$ suppression rates in the two cases is clear. We have now shown that there are types of defects produced by \={e}-irradiation, presumably Frenkel pairs of Fe vacancies and interstitials, which suppress $T_c$ at a rate much closer to the ``ideal" Abrikosov-Gor'kov rate expected for the ``symmetric model" $s_\pm$ state.  Since such a fast $T_c$ suppression due to nonmagnetic disorder cannot take place for an $s_{++}$ state, we have ruled out a non-sign changing $s-$ wave state at least for the 122 materials.  This conclusion is also consistent with the similar pairbreaking rates for magnetic and nonmagnetic impurities in Ref.~\onlinecite{Lietal2012}. It disagrees, however, with a recent detailed 5-orbital study of $T_c$ suppression by transition metal impurities \cite{Yamakawa2013} modeled by \textit{ab initio} calculations \cite{Nakamura2011} that found that $T_c$ should be suppressed to zero at a critical resistivity of $\Delta \rho_0 \sim 10~ \mu \Omega \cdot cm$, in contrast to the observed values of ($\sim 1~ m\Omega \cdot cm$) in Ref.~\onlinecite{Li2012}. We cannot reconcile the claim of $s_{++}$ pairing in Ref.~\onlinecite{Yamakawa2013} with the very small critical resistivity observed here.

One obvious concern is that this conclusion might be invalidated were electron irradiation at the energies used in this experiment to create magnetic defects, which would indeed act as strong pairbreakers in an $s_{++}$ state.  We have examined the low-temperature penetration depth data on the irradiated samples in each case for signs of a low-temperature upturn that would indicate the contribution of magnetic defects to the superfluid density \cite{Cooper1996}, and found no such terms. We are therefore confident that the evolution of the superconducting transition temperature, \tc, with the electron irradiation-induced increase of residual resistivity in \baru\ is fully consistent with the generalized treatment of the impurity scattering within the \spm\ pairing mechanism in iron-based superconductors. This conclusion is also supported by the electron irradiation study of a known two-band, but $s_{++}$ superconductor, MgB$_2$, where $T_c$ remained practically unchanged  \cite{KleinPRL2010MgB2}.

\begin{acknowledgments}

We thank SIRIUS team: B. Boizot, V. Metayer, and J. Losco for help with the electron irradiation. We also thank C. van der Beek, A. Chubukov, R. Fernandes, H. Kontani, D. Maslov, I. Mazin, F. Rullier-Albenque, T. Shibauchi, and T. Tamegai for useful discussions. The work in Ames was supported by the U.S. Department of Energy (DOE), Office of Science, Basic Energy Sciences, Materials Science and Engineering Division. Ames Laboratory is operated for the U.S. DOE by Iowa State University under contract DE-AC02-07CH11358. Work at U. Florida was partially supported by by DOE DE-FG02-05ER46236. The work at the Ecole Polytechnique was supported by EMIR network, proposal 11-11-0121. R.P. acknowledges the St. Gobain Chair position at the Ecole Polytechnique. VM acknowledges support from the Center for Emergent Superconductivity, an Energy Frontier Research Center funded by the US DOE, Office of Science, under Award No. DE-AC0298CH1088.

\end{acknowledgments}

\end{document}